# Switchable Magnonic Crystals Based on Spin Crossover/CrSBr Heterostructures

Andrei Shumilin[†], Sourav Dey[†,1], Denisa Coltuneac, Laurentiu Stoleriu and José J. Baldoví*

A. Shumilin, S. Dey, J.J. Baldoví

Instituto de Ciencia Molecular (ICMol), Universitat de Valencia, c/Catedrático José Beltrán, 2, Paterna 46980, Spain

E-mail: j.jaime.baldovi@uv.es

S. Dey

Department of Chemistry, Washington State University, Pullman, WA 99164, USA

D. Coltuneac, L. Stoleriu

Faculty of Physics, Alexandru Ioan Cuza University, Iasi 700506, Romania

Funding: European Union (ERC-2021-StG-101042680 2D-SMARTiES), Marie Curie Fellowship SpinPhononHyb2D 10110771, and RExQTCS Project (PN-IV-P6-6.1-CoEx-2024-0214) of Romanian UEFISCDI.

## Abstract

The progress of magnonics ultimately depends on material platforms that offer precise control of spin waves propagation. Here, we put forward a chemical strategy to create locally tunable magnonic crystals by integrating switchable spin-crossover (SCO) molecules with 2D van der Waals magnets. Specifically, we investigate from first principles a hybrid molecular/2D heterostructure formed by $[Fe((3,5-(CH_3)_2Pz)_3BH)_2]$ molecules (Fe-pz) deposited on a single-layer of semiconducting CrSBr. We show that Fe-pz molecules are stable on CrSBr while preserving its SCO bistability, particularly in densely packed molecular arrays. By patterning Fe-pz into periodic stripes separated by pristine CrSBr regions, the interface becomes a magnonic crystal that filters spin waves at selected frequencies. Crucially, light-driven excited spin-state trapping (LIESST) enables LS→HS switching and induces up to ~1.3% local strain in CrSBr, which in turn reshapes the magnonic band structure in a dynamic and reversible manner. These results establish Fe-pz@CrSBr as a switchable platform for on-chip, programmable magnonic devices.

[†]A. Shumilin and S. Dey contributed equally to this manuscript

# 1. Introduction

Magnonics is an emerging research field that exploits the use of magnons —quanta of spin waves— as information carriers, promising an ultra-low dissipation alternative to charge-based electronics.[1,2] A critical step toward practical magnonic technologies lies in the development of functional building blocks capable of controlling spin-wave propagation on demand. Among these, magnonic crystals (MCs) are particularly attractive due to their ability to selectively suppress spin-wave transmission at specific frequencies through periodic modulation of magnetic properties.[3] This modulation is typically achieved by combining different magnetic materials[4,5] or by locally altering their magnetic properties.[6,7] While static MCs have been studied for applications such as microwave filtering[8] and data storage,[9] the real leverage comes from switchable MCs whose magnon band structure can be reconfigured by external stimuli.[10]

Several routes toward dynamic MCs have been explored, ranging from current-generated periodic magnetic fields in meander structures[11–13] to selective magnetization reversal,[14,15] engineered magnetic textures (domains, skyrmions),[16–18] and the application of macroscopic strain.[19] However, most of these approaches rely on relatively large device architectures — typically on the order of hundreds of micrometers—which enters in conflict with a critical limitation in this field: finite propagation lengths limited by intrinsic damping. This constraint highlights the need for miniaturized, locally addressable MCs, compatible with nanoscale integration and fast actuation.[20]

Against this challenge, hybrid interfaces that combine two-dimensional (2D) van der Waals (vdW) magnets with responsive molecular materials offer a compelling path for the development of switchable MCs. On the one hand, molecular materials are compatible with surface chemistry and provide exceptional versatility and nanometric size; on the other, vdW magnets provide an atomically-thin host in which spin-wave transport can be selectively modulated via interfacial effects. Among 2D magnets, CrSBr stands out as a layered magnetic semiconductor with a relatively high Curie temperature ($T_C$ ~146 K), air-stability[21–23] and a magnon band structure that can be driven by strain.[24,25] On the molecular side, spin-crossover (SCO) compounds are key examples of bistability: sublimable Fe(II) complexes can reversibly switch between low-spin (LS) and high-spin (HS) states under temperature, pressure, or light.[26] This LS→HS transition leads to a significant elongation of the metal–ligand bonds in the HS state, driven by the occupation of antibonding orbitals, producing a local lattice expansion capable of triggering strain.[27–31] Although equilibrium SCO transition typically occur at ~150–300 K –above the temperature window usually considered for CrSBr magnonics– the so-called light-induced excited spin state trapping (LIESST) effect enables LS→HS conversion via photon absorption even at temperatures of 10 K or below.[32–34] Despite their potential, the integration of SCO molecules into solid-state nanodevices remains challenging primarily because deposition can degrade molecules and interfacial interactions may quench bistability. A particular robust candidate is [Fe((3,5-$(CH_3)_2$Pz)$_3$BH)$_2$] (Pz = pyrazolyl, hereafter Fe-pz), which has shown remarkable stability on a variety of surfaces, while retaining SCO behavior even under ambient conditions.[35,36]

In this work, we employ a multiscale computational strategy to connect molecular switching with magnon transport in Fe-pz@CrSBr. Through first-principles calculations, we investigate the stability of the adsorbed molecule, interfacial electronic and magnetic effects, and the preservation of SCO bistability, including cooperative effects in densely packed molecular assemblies. Then, we develop an effective elastic model that captures non-uniform strain fields over length scales up to ~100 nm. Finally, we translate these strain landscapes into local modification of magnon transport in CrSBr, using a combination of spin Hamiltonian, linear spin-wave theory and micromagnetic simulations. Our results establish Fe-pz@CrSBr as a stimuli-responsive, switchable MC, capable of reversible modulation of magnon propagation via molecular spin switching.

## 2. Results and discussion

### 2.1. Fe-Pz molecule deposition on CrSBr

CrSBr crystallizes in an orthorhombic structure with *Pmmn* symmetry in the framework of a rectangular unit cell. Our simulations, at the monolayer limit, result in optimized lattice parameters of $a$ = 3.589 Å and $b$ = 4.792 Å, being in good agreement with previous studies.[24] Each unit cell contains two magnetic Cr atoms with S=3/2, exhibiting long-range ferromagnetic ordering. As SCO counterpart, we use Fe-pz molecule, which possesses a Fe(II) center coordinated by two pyrazolylborate ligands, resulting in a distorted FeN$_6$ octahedral geometry. The average Fe–N bond lengths are found to be 1.995 Å and 2.178 Å for the low spin (LS) and high spin (HS) states, respectively.

To investigate the effect of surface adsorption of Fe-pz molecules on the 2D magnet, we begin by constructing a hybrid molecular/2D heterostructure consisting of a single Fe-pz deposited in a 6 × 4 supercell of CrSBr (**Figure 1a**). This setup simulates a sparse coverage of CrSBr monolayer and ensures negligible interaction between neighboring molecules. We consider four adsorption geometries, placing the magnetic center (Fe) above Br, Cr, S, and hollow sites (Figure 1b), in order to obtain the most stable molecular configuration after adsorption. We also vary the angle θ between the B–Fe–B axis and the CrSBr plane from 10° to 68° (Figure S6 in ESI). The calculated equilibrium distances between Fe-pz and CrSBr range from 2.80 to 3.00 Å, depending on orientation, consistent with a physisorption process. Among all configurations, the most stable geometry is the one that has the Fe atom above the hollow site with θ = 23.7°. Nevertheless, the energy differences with the other tested configurations are relatively small (Tables S3 and S4).

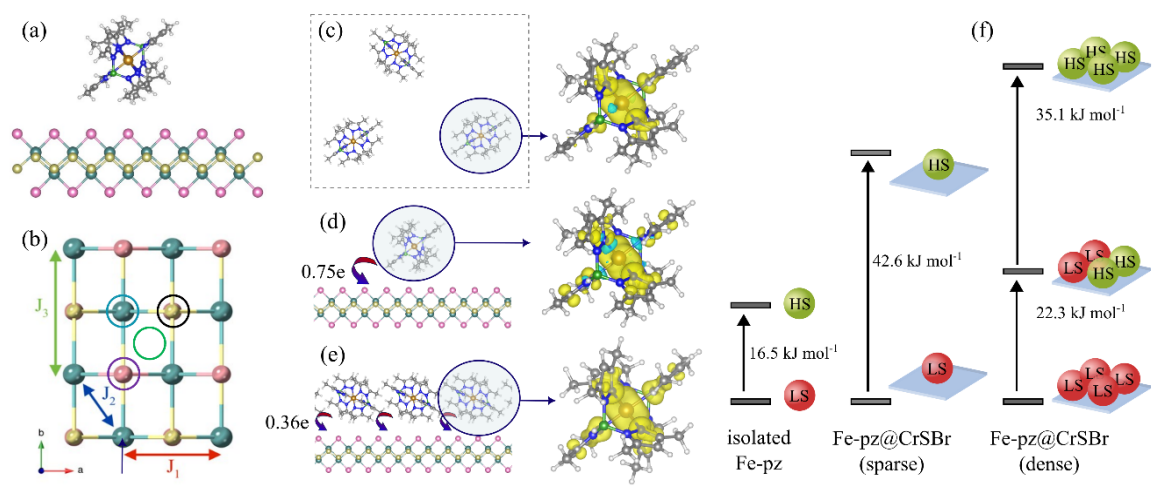

**Figure 1.** (a) Lateral view of a hybrid molecular/2D heterostructure formed by Fe-pz adsorbed on single-layer CrSBr. Colour code: Cr (cyan), Br (pink), S (yellow), Fe (brown), B (green), N (blue), C(silver), H (white). (b) Top view of CrSBr monolayer showing four adsorption sites (top of: Br (purple circle), Cr (cyan circle), S (black circle), hollow (green circle)) and exchange interactions $J_1$-$J_3$ in CrSBr. (c-e) Spin density in Fe-pz molecule in different configurations: (c) in gas phase, (d) single molecule deposited on CrSBr and (e) dense array of the molecules deposited on CrSBr. The charge transfer from Fe-pz to CrSBr is shown by arrows. (f) the energies of LS→HS transitions of a Fe-Pz molecule in different configurations, calculated per molecules changing their state.

Our calculations show that adsorption induces minor modifications in the LS state under sparse coverage but has a stronger effect on the HS state. This leads to an increase in the spin transition energy from 16.5 kJ mol$^{-1}$ for the gas-phase molecule to 42.6 kJ mol$^{-1}$ under sparse coverage.

These modifications include a notable elongation of the Fe–N bond in the HS state (~0.11 Å compared to ~0.01 Å in LS) and larger variations in the energy splitting between the barycenters of the $t_{2g}$ and $e_g$ orbitals, as calculated with ab initio ligand field theory (AILFT). Consequently, the adsorption energies differ for LS (–0.82 eV) and HS (–0.52 eV) states (see Section 4 of SI for additional details). The stronger interaction in the LS state is attributed to its more spatially extended orbitals (Figure S10), which enhance overlap with the substrate, consistent with previous studies of Fe-pz on Au surfaces.[35]

Interestingly, our results indicate that the difference in molecular modifications between LS and HS states diminishes at higher molecular coverages due to molecule–molecule interactions. To illustrate this, we design a dense coverage setup with four Fe-pz molecules in a single 6 × 4 CrSBr supercell.[35,36] Three spin configurations were explored: HSHSHSHS (all HS), HSHSLSLS (two HS, two LS), and LSLSLSLS (all LS). After structural optimization, molecules remain at similar equilibrium positions with θ~19° (See Table S5 in SI for the details). Figure 1f compares spin transition energies for (i) isolated molecules, (ii) a single molecule adsorbed on CrSBr, and (iii) four molecules adsorbed on CrSBr. For dense coverage, the spin transition energy decreases significantly: 22.3 kJ mol$^{-1}$ for LSLSLSLS → HSHSHSHS and 35.1 kJ mol$^{-1}$ for HSHSLSLS → HSHSHSHS (accounting for molecules switching their spin states). These values are close to the estimated spin transition energy of bulk Fe-pz molecular crystals (29.5 kJ mol$^{-1}$) [42], indicating that cooperative effects are crucial for preserving SCO properties on CrSBr surfaces, whereas sparse coverages partially suppress these properties.

To further probe this effect, we performed Bader charge analysis and examined spin densities for molecules in the HS state under sparse and dense coverage (Figure 1c–e). Charge transfer per molecule is 0.36 e under dense coverage, compared to 0.75 e for sparse coverage. In the sparse case, charge transfer is accompanied by significant spin density modification (Figure 1d) and a substantial reduction in Fe magnetization by 0.8 $\mu_B$, suggesting partial Fe(II) → Fe(III) oxidation upon adsorption. This accounts for the approximately fourfold increase in spin-transition energy. Conversely, molecule–molecule interactions in dense coverage restore spin densities resembling those of gas-phase Fe-pz (Figure 1c,e), with total magnetization of 3.7 $\mu_B$ per molecule. Additional evidence for Fe oxidation in sparse coverage and restoration to Fe(II) in dense coverage, based on average Fe–N bond lengths and AILFT analysis, is provided in Tables S1 and S2 of SI.

## 2.2. Magnetic properties of CrSBr@Fe-pz heterostructures

To investigate how the deposition of Fe-pz molecules—and their switching between LS and HS states— affects the magnetic and magnonic properties of CrSBr, we map the total energies from DFT calculations with four different spin configurations (one ferromagnetic and three antiferromagnetic) onto a Heisenberg spin Hamiltonian commonly used to describe CrSBr:

$$\widehat{H} = \sum_i \sum_{n=1}^{3} J_n \mathbf{S}_i \mathbf{S}_{j(n,i)} \qquad (1)$$

Here index *i* enumerates the Cr spins, and n =1, 2 or 3 corresponds to the first-, second- and third-neighbor magnetic exchange interactions, respectively, as shown in Figure 1b. The details of the mapping are given in Section 2.6 of SI. Although Fe-pz molecules in the HS state carry spin, their exchange coupling with Cr atoms is found to be negligible and is thus not included in the Hamiltonian. The computed values of the exchange parameters $J_1$-$J_3$ are listed in **Table 1**.

**Table 1.** Computed magnetic exchange (in meV) of pristine CrSBr, sparse and dense Fe-pz covering of CrSBr.

|  | Pristine CrSBr | sparse Fe-pz@CrSBr | | dense Fe-pz@CrSBr | |
|---|---|---|---|---|---|
|  |  | HS | LS | HSHSHSHS | LSLSLSLS |
| $J_1$ | 3.637 | 3.610 | 3.847 | 3.870 | 3.853 |
| $J_2$ | 4.015 | 4.134 | 4.223 | 4.114 | 4.117 |
| $J_3$ | 2.700 | 3.336 | 3.448 | 4.665 | 4.798 |

Our results show that the magnetic exchange interactions in CrSBr are modified upon Fe-pz adsorption, with the most significant change observed in $J_3$, which increases from 2.7 meV (in pristine material) to 3.4 meV (sparse Fe-pz@CrSBr) and further to 4.7 meV (dense Fe-pz@CrSBr). However, even in the dense molecular covering, the influence of Fe-pz transitions between LS and HS states on CrSBr magnetic properties remains small. It is important to emphasize that these configurations—with one or four Fe-pz molecules per supercell—do not form a continuous molecular layer but rather represent isolated molecules adsorbed on CrSBr. Their spatial arrangement follows the rectangular lattice of the CrSBr monolayer, and the intermolecular distances (9.6 Å along the *a*-axis and 10.8 Å along the *b*-axis, measured center-to-center) are larger than those in Fe-pz crystals (8.6 Å in the LS state and 9.0 Å in the HS state).[37] Here is important to note that due to the lattice mismatch between CrSBr and Fe-pz, it is computationally unfeasible to model a true CrSBr@Fe-pz bilayer using ab initio methods within reasonably sized supercells.

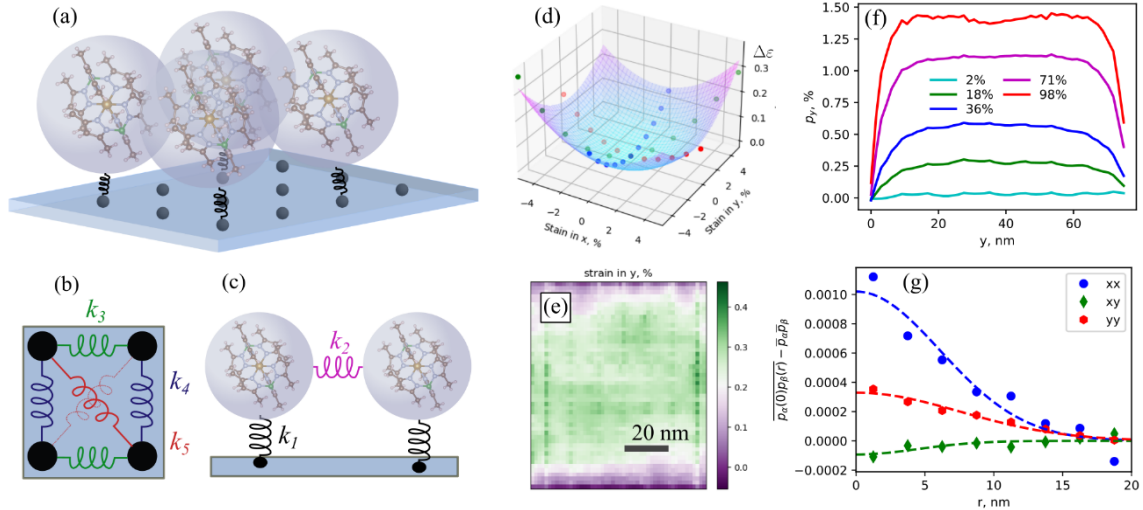

**Figure 2.** (a) Schematic view of the elastic model. (b) Springs describing elastic interactions in a CrSBr monolayer. (c) Springs describing molecules and their interaction with CrSBr. (d) *ab initio* calculated elastic energies compared with Harmonic approximation. (e) Spatial dependence of strain $p_y$ in a mixed spin state of the molecules (with 71% of molecules in HS state). (f) Averaged strain $p_y$ as a function of $y$ for 2, 18, 36, 71, 98% of molecules in the HS state. (g) Correlation functions of the quasi-random strains $p_x$ and $p_y$ in the mixed spin state (71% of the molecules in HS state).

Then, we investigate the effect of mechanical strain on the magnetic properties of CrSBr due to Fe-pz switching LS→HS transitions. For that, we develop an elastic framework inspired by the approach in Ref.[37] (see **Figure 2**a). In this framework, Fe-pz molecules are treated as rigid spheres whose radius increases from 3.0 Å to 3.3 Å during the LS→HS transition. Each molecule interacts with the CrSBr surface via an effective spring connecting it to its preferred adsorption site. Based on first-principles calculations while varying the molecule–substrate distance, we estimate the equilibrium spring length to be ~3 Å (from the sphere's surface) with a spring constant $k_1$=0.45eV Å$^{-2}$ (see the details in Section 2.7 of SI). Following Ref.[37], we describe the Fe-pz layer as a hexagonal array of spheres coupled by springs of stiffness $k_2$ = 1.4eV Å$^{-2}$, with an equilibrium surface-to-surface distance of 3 Å (Figure 2c). To characterize the elasticity of the CrSBr substrate, we perform *ab initio* calculations under uniaxial and biaxial strains ranging from –4% to +4% along the *a* and *b* directions. As shown in Figure 2d, for strains below 3%, the energy–strain relationship remains in the harmonic regime. Accordingly, we describe CrSBr as a 2D network of zero-radius nodes representing the optimal adsorption sites of Fe-pz. These nodes are linked by springs $k_3$-$k_5$ (Figure 2b), with parameters chosen to reproduce the macroscopic elasticity tensor of CrSBr. Further modelling details are provided in Section 3 of SI. This elastic model enables the identification of mesoscopic strains induced in CrSBr by the monolayer of Fe-pz molecules. We model an 80×80nm square sample of CrSBr interacting with a monolayer of Fe-pz with *x* and *y*-direction in the simulation corresponding to *a* and *b*-directions of CrSBr. Adsorption of the Fe-pz molecules—assumed to be in the low-spin (LS) state—always introduces some degree of strain due to the lattice mismatch between CrSBr and the Fe-pz molecular crystal. However, this strain is generally minor (less than 0.5%) and does not significantly affect the magnetic properties of the system. The strain increases as some molecules transition to the HS state via the LIESST effect, placing the system in a mixed-spin state, and reaches its maximum when all molecules are in the HS state. The typical non-uniform strain induced by the adsorbed molecules is illustrated in Figure 2e, which shows the strain component along the *y*-direction in the mixed-spin state. As expected under open boundary conditions, the strain vanishes at the top and bottom edges ($y=0$ and $y=80$ nm), but not at the left and right edges, where the system remains free to deform. To better analyze the strain profile, we average the y-direction strain, over the *x*-coordinate and plot it as a function of *y* (see Figure 2f). While the average strain in the LS state is minimal, it increases significantly in the HS state, reaching values close to 1.3%. Mixed-spin states produce intermediate strain levels, consistent with a partial transition of the molecules to HS-state. In addition to the smooth, averaged strain profile, Figure 2e also reveals a quasi-random strain component. This arises from two main factors: the intrinsic lattice mismatch between CrSBr and Fe-pz, and the random spatial distribution of HS molecules in the mixed-spin state. In this work, we treat this fluctuating strain as a random variable, characterized by a distance-dependent correlation function, which is presented in Figure 2g. More information on the treatment of the non-uniform strain spatial distribution maps can be found in Section 4 of SI.

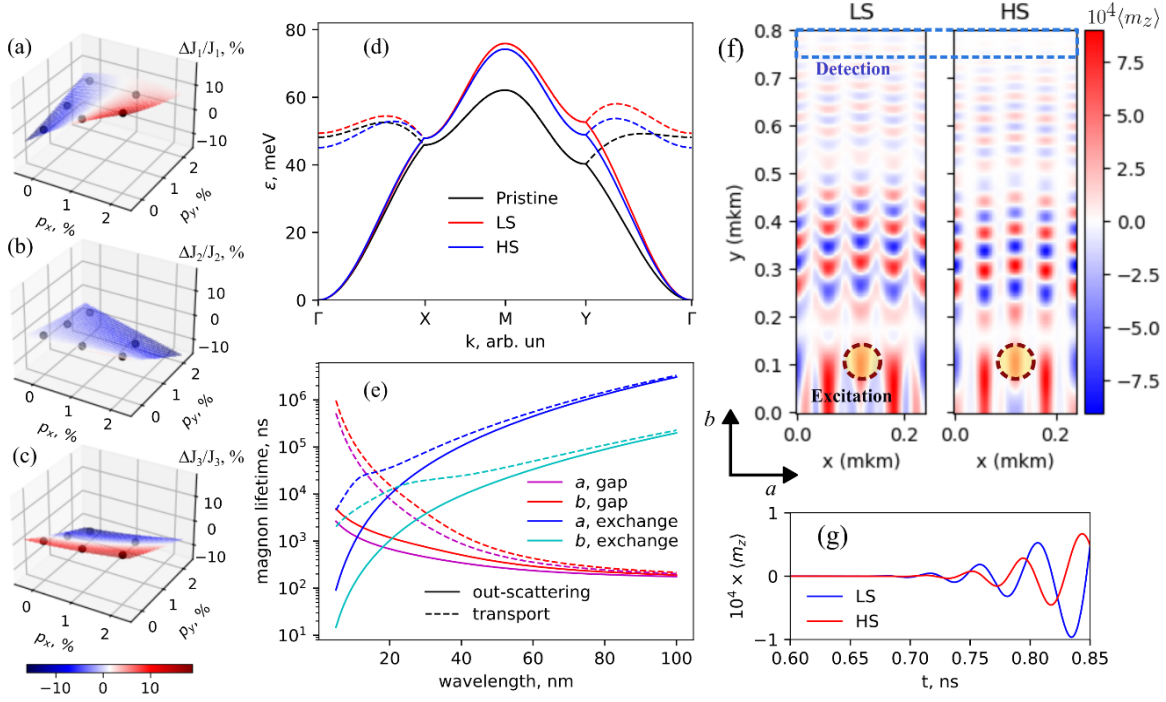

**Figure 3.** (a)-(c): The relative modification of the exchange interactions $J_1$-$J_3$ respectively due to the applied strain. Black points correspond to ab initio simulation while the surfaces show its interpolation to the arbitrary strains. (d) The magnon spectra of the pristine material, and material with molecules in LS/HS states with the averaged induced strain. (e) The magnon lifetime due to the scattering on the correlated disorder in the intermediate spin state. (f) Spin wave propagation in the strips of SCO/CrSBr after the initial excitation at the bottom side of the strip. (f) The averaged x-magnetization of the top area of the strip as a function of time.

Because the calculated strain profiles vary smoothly over ~10 nm, we use an adiabatic approximation to analyze the impact of the SCO molecule switching on magnon properties. This assumes the magnon dispersion evolves slowly in space and can be locally treated as if under uniform strain. To quantify strain effects, we extracted the strain dependence of exchange interactions $J_1$–$J_3$ by interpolating *ab initio* results at six strain values (**Figure 3**a–c). Based on these, we calculated the magnon spectra for pristine CrSBr and CrSBr@Fe-pz heterostructures in both LS and HS molecular states, including strain effects (Figure 3d). While the overall LS→HS difference is moderate, the effect is notably stronger for magnons propagating along the *b*-axis (Γ–Y direction), due to the strong strain sensitivity of $J_3$. Along the *a*-axis, competing effects from $J_1$ and $J_2$ largely cancel, leading to minimal strain influence on the corresponding magnon modes.

Although the differences in magnon spectra between LS and HS states appear modest, they can significantly affect time- and space-resolved magnon transport experiments. One has to realize that even slight variations in magnon group velocity accumulate over distance, producing measurable effects. To illustrate this, we simulate magnon propagation in a 0.24 × 0.8 μm strip of single-layer CrSBr by solving the Landau-Lifshitz-Gilbert (LLG) equations. Magnons are emitted in an ~80 nm region, mimicking an excitation from a spin-torque nano-oscillator[38,39]. This generates a 30 ps field-like torque pulse, launching a coherent magnon wave packet with a broad wavelength range. Since group velocity is inversely proportional to wavelength, short-wavelength magnons travel faster and reach farther at a given time delay (Figure 3f). Notably, group velocities are higher across all wavelengths in the LS state than in the HS state. As a result, the magnon pulse detected far from the excitation site differs in both shape and amplitude depending on the molecular spin state. Figure 3g shows the time-dependent out-of-plane magnetization component $m_z$, averaged over a small region near the strip's far end (highlighted in Figure 3f). The signal remains negligible before 0.75 ns, then rises sharply. Between 0.75 and 0.8 ns, the $m_z$ signal is significantly stronger in the LS state and exhibits a phase shift relative to the HS case.

While the LS→HS transition primarily alters magnon velocities, the presence of a mixed-spin state introduces a magnon decay due to correlated pseudo-disorder in the strain field. To investigate this, we analyze the influence of correlated disorder in the Heisenberg exchange interactions and in the magnon gap, which usually arises from anisotropic exchange. Although our *ab initio* calculations capture only the Heisenberg component, we estimate the strain dependence of the magnon gap using reported results for pristine CrSBr from the literature[24]. Both contributions, namely exchange disorder and gap fluctuations, are treated within the Boltzmann equation formalism using the Born approximation, as described in detail in section 5 of SI. Our findings indicate that magnon relaxation in this system cannot be adequately described by conventional Gilbert damping. This is because the relaxation rate is not proportional to the magnon frequency. Instead, we present the magnon out-scattering and transport relaxation times as functions of magnon wavelength in Figure 3e. These contributions are also analyzed using a simplified analytical toy model (see Eqs. S.5.13 and S5.14). Both the mechanisms are strongly enhanced by increasing correlation length of disorder. The out-scattering due to Heisenberg exchange disorder is strong for the short-wavelength magnon reaching 15ns for 5nm magnons propagating along b-direction. However, it is suppressed by a huge factor $(a_J/\lambda)^4$ when the magnon wavelength λ becomes larger than typical inter-atomic distance $a_J$. At these wavelengths the correlated fluctuations of the small magnon gap become increasingly important in agreement with previous studies of magnon scattering by phonons.[40] This results in magnon relaxation times ~180 ns.

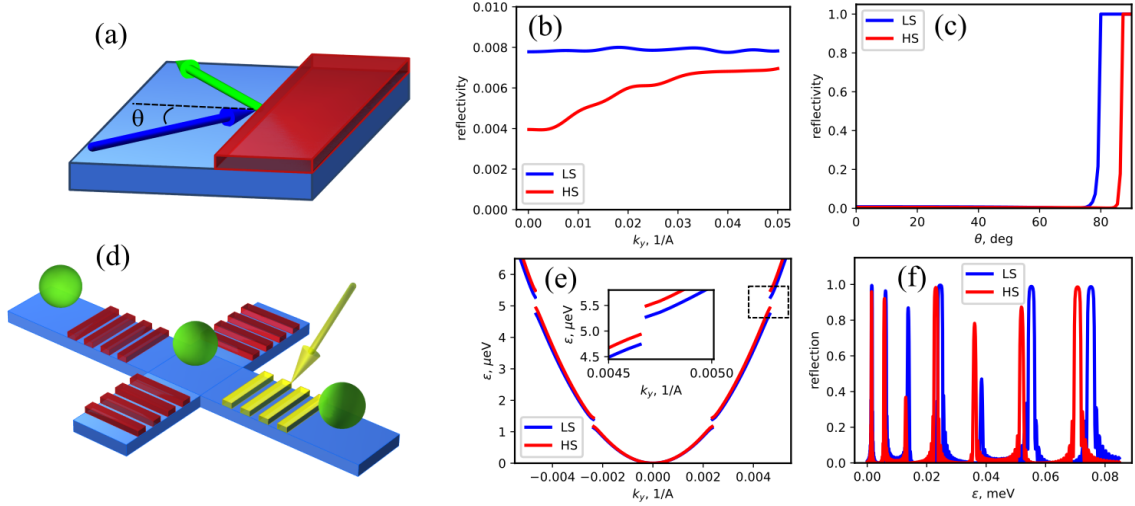

**Figure 4.** (a) Sketch of magnon reflecting from CrSBr:SCO. (b) magnon reflectivity for normal propagation as a function of wavevector. (c) Magnon reflectivity dependence on the incidence angle for the absolute value of wavevector equal to $0.01$Å$^{-1}$. (d) Sketch of the switchable MC applied as a tunable gate between qubits. (e) Acoustic magnon dispersion in MC, inset shows the MC magnon gap in close scale. (f) The magnon reflectivity in a finite MC containing 20 periods of 80nm of Fe-pz@CrSBr and 45nm of pristine CrSBr.

## 2.3. CrSBr@Fe-pz as a platform for magnonics

The contrast between pristine CrSBr and the regions in the 2D material covered with SCO molecules enables the design of magnonic devices, where molecular switching via LIESST offers a control mechanism. To explore this, we calculate the reflectivity of a monochromatic spin wave at the interface between pristine CrSBr and a large area covered with Fe-pz molecules (see **Figure 4**a). Details of the calculations are described in Section 6 of SI. Figure 4b shows the reflectivity of an acoustic magnon propagating along the *b*-axis at normal incidence. Although the overall reflectivity is low (~1%), the difference between LS and HS states is substantial. This demonstrates that spin-state switching can, in principle, modulate magnon transport. Figure 4c presents the angular dependence of reflectivity for a fixed wavevector magnitude $k=0.01$Å$^{-1}$ as a function of incidence angle θ. The reflectivity approaches unity at for θ>80° for LS state of molecules and at θ>89° for HS state of the molecules, due to the onset of total internal reflection.

The reflectivity can be significantly enhanced –though only for specific magnon wavelengths– by introducing a MC composed of alternating stripes of pristine CrSBr and CrSBr@Fe-pz heterostructure, i.e. by the creation of molecular stripes on the surface of the 2D magnet. Figure 4e illustrates the magnon dispersion in an idealized (infinite) MC with zero damping which consists of 80 nm-wide stripes of CrSBr@Fe-pz separated by 45 nm-wide regions of pristine CrSBr. While the overall dispersion resembles that of pristine CrSBr, magnon band gaps emerge at specific wavevectors due to destructive interference, which inhibits magnon propagation. Crucially, the position and width of these band gaps depend on the spin state of the Fe-pz molecules, which allows molecular control of the magnon spectra. While in the idealized MC, magnons with energies falling within the band gaps would be completely reflected at the boundary, in finite systems they would sometimes tunnel through MC, leading to high—but not perfect—reflectivity. Devices based on this principle are known as *Bragg mirrors*, which are well-established components in both classical and quantum optics[41,42]. Figure 4f shows the computed reflectivity of a magnon wave incident on a structure consisting of 20 periods of the magnonic crystal. A Gilbert damping factor of α=0.001 is assumed, which is a typical value for Cr-based vdW magnetic materials.[43–46] The reflectivity exceeds 95% for certain magnon energies, which differ depending on whether the Fe-pz molecules are in the LS or HS state. For instance, at a magnon energy of 0.075 meV, the reflectivity is approximately 98% in the LS state but drops to around 1% in the HS state. This clearly demonstrates how hybrid SCO/CrSBr heterostructures can function as *switchable Bragg mirrors* for magnonic applications.

According to our results, we envision that switchable MCs based on SCO/2D magnetic heterostructures could serve as key components for the integration of 2D magnets into quantum computing architectures. An important requirement for efficient quantum computations is the ability to controllably switch interactions between qubits. In magnon-based platforms, qubits can be designed to be in resonance with magnons with specific frequencies. By placing a magnonic Bragg mirror between two qubits, one can effectively control their coupling: in the "closed" state, the mirror reflects most of the magnonic signal, preventing interaction. However, when an external signal triggers the LS→HS transition in the SCO molecules, the magnonic band structure of the Bragg mirror shifts, thus tuning the position of the band gaps (see Figure 5d). This transition can open a transmission window for magnons that match the qubit resonance frequency, thereby "opening" the channel and enabling qubit-qubit coupling. Such functionality provides a pathway for dynamic, on-chip control of qubit connectivity, paving the way for programmable magnonic quantum circuits.

## 3. Conclusion

In summary, we introduce a molecular route to switchable magnonic crystals by integrating spin-crossover molecules with 2D van der Waals magnets. Our first-principles calculations show that Fe-pz molecules remain stable upon deposition on CrSBr while preserving their bistability, especially in densely packed arrays, and that cooperative packing constraints can favor mixed-spin textures. We identify strain as the key transduction channel from molecular switching to magnon transport. Indeed, lattice mismatch induces a modest distortion in the LS state that increases upon LS→HS conversion, generating a controllable strain landscape. Importantly, these strain effects directly impact magnon propagation in CrSBr: (i) correlated strain disorder enhances magnon scattering in the mixed spin state, whereas (ii) the magnon group velocities and the probability of magnon reflection at the interface between pristine CrSBr and Fe-pz covered regions are shifted when the full switching to the HS state occurs.

Finally, we demonstrate that periodic Fe-pz stripes patterned on CrSBr operate as a switchable magnonic crystal/Bragg mirror, with spin-state-dependent band gaps and high-reflectivity windows. This provides a chemically addressable framework to gate magnon transmission at selected frequencies, opening a viable route toward the dynamic control of this next-generation magnetic devices, paving the way for on-chip programmable magnonics.

## 4. Methods

The zero-point energy (ZPE) of the HS and LS molecules was calculated from the geometry optimization in the gas phase. The calculations were performed using B3LYP functional using Gaussian09 program package.[47] We have used Ahlrichs triple-$\zeta$ plus polarisation basis set for Fe and double-$\zeta$ plus polarisation basis set for remaining atoms[48]. After obtaining the frequencies of the 3N-6 vibrational modes of the molecules, the ZPE of each spin state was calculated as

$$ZPE_{HS/LS} = \sum_{i=1}^{Nvib} \frac{1}{2} h\nu_i$$

The difference between ZPEs of the LS and HS states ($\Delta ZPE = ZPE_{HS} - ZPE_{LS}$) is found be 10.6 kJ mol$^{-1}$, therefore ZPE energy tries to stabilize the HS state.

We have performed CASSCF/NEVPT2 calculations on Fe-Pz molecule to estimate the ligand field strength of the metal center for each spin state which was determined from ab initio ligand field theory (AILFT) approach as implemented in ORCA program package[49]. The relativistic effect in our calculations was taken into account by Douglas-Kroll-Hess (DKH) Hamiltonian. All the calculations were performed with def2-SVP basis set apart from N and Fe for which def2-TZVP(-f) and def2-TZVP basis sets were employed, respectively. We have selected 6 electrons for the five 3d orbitals in the active space for state-average CASSCF calculations. With this active space, we have computed the energies of 5 quintets, 45 triplets and 50 singlets.

We have investigated the deposition of Fe–pz molecules on CrSBr by means of periodic density functional theory calculations, using VASP (Vienna ab initio simulation package) program package in the framework of projected-augmented wave (PAW) method[50]. We have employed revised Perdew–Burke–Ernzerhof (rPBE) functional in all our calculations. This exchange-correlation functional chosen as the benchmark calculations[36] concluded that HS-LS energy gap estimated from this functional is in excellent agreement with experiment compared to generalized gradient approximation (GGA) functionals where the spin transition energy strongly depends on the Hubbard U parameters. Furthermore, Calzado and co-workers demonstrated that rPBE functional provides an estimate of HS-LS energy gap that is in excellent agreement with those obtained from the wave function-based calculations such as CASSCF/NEVPT2 and CASPT2.[35]

We have employed an energy cutoff of 500 eV for representing the valence electrons of the plane-wave basis set with a Γ-point Brillouin zone. The lattice parameters of CrSBr monolayer were optimized and the resulting lattice parameters ($a$ = 3.589 Å and $b$ = 4.792 Å) were used to build the 6 ×4 supercell in modelling the hybrid heterostructure. We have considered more than 18 Å vacuum in the $z$ direction to prevent interaction in this direction. During the calculations, the electronic relaxations were performed until the change in total energy between two consecutive steps <10$^{-6}$ eV and ionic relaxations were performed until the Hellmann−Feynman forces were < 0.025 eV Å$^{-1}$. To obtain different percentage or fractions of HS molecules on the monolayer, the NUPDOWN option was employed which is 0 for the pure LS phase, 8 with ½ of HS molecules and 16 for the pure HS phase. After the optimization, we removed the NUPDOWN restriction and total energy of each phase was recalculated. The adsorption energy ($E_{ads}$) was calculated with respect to isolated molecule and 2D CrSBr monolayer as: $E_{ads} = E_{hetrostructure} - E_{CrSBr} - E_{molecule}$. Therefore, negative adsorption energy corresponds to favorable adsorption.

The mechanical properties of the coupling between the Fe-pz and CrSBr layers were studied using a mechanoelastic model[51–53]. The system was modeled as two layers: the top layer comprised 9900 spheres arranged in a triangular lattice representing Fe-pz molecules, while the bottom layer contained 36516 spheres arranged in a rectangular grid, representing Fe-pz adsorption sites in the CrSBr layer. Fe-pz spheres were assigned radii of 3.05 Å in the low-spin (LS) state and 3.25 Å in the high-spin (HS) state. Spheres in the CrSBr layer had a fixed radius of 0.1 Å. Elastic interactions were modeled using springs with varying constants. Fe-pz molecules were connected by springs with an elastic constant of 1.38 eV Å$^{-2}$. In the CrSBr layer, spring constants were 8.134 eV Å$^{-2}$ and 3.695 eV Å$^{-2}$ along the *x* and *y* directions, respectively, and 0.95 eV Å$^{-2}$ along diagonals. The Fe-pz layer was positioned 3 Å above the CrSBr layer, with each Fe-pz sphere connected to its nearest CrSBr counterpart by a spring of 1.38 eV Å$^{-2}$. The simulation proceeded in discrete steps, randomly switching a subset of Fe-pz molecules to the HS state and then solving the equations of motion to reach a new equilibrium, allowing in-plane movement in both layers to preserve planarity. Each step yielded updated equilibrium positions and corresponding strain distributions.

The magnon spectrum for pristine CrSBr and CrSBr@FePz at different strains was calculated by interpolating the parameters $J_1$-$J_3$ of model spin Hamiltonian and applying rad-tools software for the calculation of magnon distribution.[54] The resulting distribution are interpolated by a 2-nd order polynomial function for the description of acoustic mangons with LLG, Boltzmann equation and the linear spin-wave theory. LLG simulations were performed with MuMax3 program package,[55] by combing the uniform exchange stiffness of 4 meV A$^{-1}$ with an additional custom filed tuned to represent realistic anisotropic exchange stiffness in CrSBr@FePz derived from the low-energy interpolation of the calculated magnon spectra. The lattice with 120×400 nodes in real space and a timestep of $10^{-14}$ s were used to integrate LLG. The results are tested for the robustness against these parameters.

Magnon relaxation time in the mixed spin states is determined by (i) calculating the spatial distributions of the Hamiltonian parameters $J_1$-$J_3$ and magnon gap in the different quasi-random configuration, (ii) reducing the distributions to the set of disorder correlation function dependent only on the distance between two points which are approximated as gaussian curves. Finally, magnon relaxation due to the correlated disorder is calculated with (iii) a Boltzmann equation within the Born approximation[56,57]. Linear spin-wave theory is applied by considering the strain in *y*-direction to be averaged over *x*-direction and introducing the natural boundary condition between the regions with different strain.[58] The spacers of pristine CrSBr between strips of CrSBr@FePz are considered to have zero strain in *y*-direction and the strain in *x*-direction equal to the average strain in CrSBr@FePz strips.

**Acknowledgements**

The authors acknowledge the financial support from the European Union (ERC-2021-StG-101042680 2D-SMARTiES), the Marie Curie Fellowship SpinPhononHyb2D 101107713, the Spanish Government MCIU (PID2024-162182NA-I00 2D-MAGIC), the María de Maeztu Centre of Excellence Program CEX2024-001467-M funded by MICIU/AEI/10.13039/501100011033, the Generalitat Valenciana (grant CIDEXG/2023/1) and RExQTCS Project (PN-IV-P6-6.1-CoEx-2024-0214) of Romanian UEFISCDI. The calculations were performed on the HAWK cluster of the 2D Smart Materials Lab hosted by Servei d'Informàtica of the Universitat de València.

**Conflict of interest.**

The authors declare no conflict of interest.

**Data Availability Statement**

All data supporting the findings of this study are available within the article and/or the SI Appendix. The raw data are available from the corresponding author upon reasonable request.

**Supporting Information**

The Supporting Information provides detailed descriptions of the numerical and analytical methods used, including *ab initio* simulations, the elastic model, the Boltzmann formalism, and linear spin wave theory, along with comprehensive results.